\documentclass[letter,a4paper,fleqn,usenatbib]{mnras}

\usepackage{newtxtext,newtxmath}

\usepackage[T1]{fontenc}
\usepackage{ae,aecompl}
\usepackage{float}

\usepackage{graphicx}	
\usepackage{amsmath}	
\usepackage{amssymb}	

\def\sn{\hbox{S/N}}  
  
\def\vsin{\hbox{$v \sin i$}}  
  
\def\kms{\hbox{km\,s$^{-1}$}}

\def\degr{\hbox{$^\circ$}}

\def\kis{\hbox{$\chi^2$}}

\title[Spot distribution and evolution on Vega]{Spot distribution and fast surface evolution on Vega}

\author[P. Petit et al.]{
P. Petit,$^{1}$\thanks{E-mail: ppetit@irap.omp.eu (PP)}
E.M. H\'ebrard,$^{3}$
T. B\"ohm$^{1}$
C.P. Folsom$^{1}$
and F. Ligni\`eres$^{1}$
\\
$^{1}$IRAP (Institut de Recherche en Astrophysique et Plan\'etologie), Universit\'e de Toulouse, CNRS, CNES, UPS, Toulouse, France\\
$^{3}$Department of Physics and Astronomy, York University, Toronto, ON M3J 1P3, Canada
}

\date{Accepted ??. Received ??; in original form ??}

\pubyear{2017}

\begin{document}
\label{firstpage}
\pagerange{\pageref{firstpage}--\pageref{lastpage}}
\maketitle

\begin{abstract}
Spectral signatures of surface spots were recently discovered from high cadence observations of the A star Vega. We aim at constraining the surface distribution of these photospheric inhomogeneities, and investigating a possible short term evolution of the spot pattern. Using data collected over five consecutive nights, we employ the Doppler Imaging method to reconstruct three different maps of the stellar surface, from three consecutive subsets of the whole time-series. The surface maps display a complex distribution of dark and bright spots, covering most of the visible fraction of the stellar surface. A number of surface features are consistently recovered in all three maps, but other features seem to evolve over the time span of observations, suggesting that fast changes can affect the surface of Vega within a few days at most. The short-term evolution is observed as emergence or disappearance of individual spots, and may also show up as zonal flows, with low-latitude and high latitude belts rotating faster than intermediate latitudes. It is tempting to relate the surface brightness activity to the complex magnetic field topology previously reconstructed for Vega, although strictly simultaneous brightness and magnetic maps will be necessary to assess this potential link.
\end{abstract}




\section{Introduction}

During the last few years, observations of magnetic fields in intermediate mass stars achieved a number of significant advances. During the previous decades, studies following the pioneering work of \cite{babcock47} have accumulated evidence that tepid stars displaying Ap/Bp chemical peculiarities host strong, stable, and simply shaped magnetic fields. This well-documented population represents about 10\% of all A and B stars (e.g. \citealt{wolff68}), and generally displays polar field strengths from a few hundred gauss \citep{auriere07} to a few tens of kilo gauss (e.g. \citealt{mathys17}). Recent studies challenge the lower magnetic threshold proposed by Auri\`ere et al. (e.g. \citealt{fossati15, alecian16, blazere16b}), and sometimes highlight a field geometry significantly more complex than a simple dipole (e.g. \citealt{kochukhov04}).

Beside this group of stars, magnetic fields about two orders of magnitudes weaker were discovered in the normal A star Vega \citep{lignieres09}, and in several bright Am stars \citep{petit11,blazere16a}. The repeated detection of weak magnetic fields in a small sample of A stars is potentially indicative of a widespread phenomenon in intermediate mass stars, an hypothesis that gains support from the frequent detection of rotational modulation in main sequence stars of the same mass range \citep{balona13}. The physical origin of this magnetism is still a subject of debate (see $e.g.$ \citealt{braithwaite13} or \citealt{gaurat15} for proposals in this domain, or \citealt{braithwaite15} for a review).

Vega is a rapidly rotating A0 star seen with a low inclination angle \citep{aufdenberg06, takeda08}, with chemical abundances characterized by a $\lambda$~Bo\"otis type metal depletion \citep{venn90}. Among the small number of stars investigated so far by means of deep polarimetric observations, Vega has benefited from a number of complementary studies. Following the initial Zeeman detection, the magnetic geometry of the star was reconstructed using the Zeeman-Doppler Imaging tomographic method (ZDI, \citealt{semel89,donati06}), revealing a region of radially-oriented magnetic field near the rotation pole, and several weaker magnetic spots at lower latitudes \citep{petit10}. The best ZDI model was obtained for a rotation period of 0.72~d, a value later revised to 0.678~d \citep{alina12}. From an intensive observing campaign spread over five consecutive nights, \cite{boehm15} concluded that several spectral features were modulated with a period of 0.678~d, and detected unambiguous signatures of surface spots in line profiles. 

We propose here to go one step further in the analysis of the observations of \cite{boehm15}, by using their time-series to reconstruct the distribution of surface inhomogeneities on Vega, with a tomographic approach.

\section{Observations, multi-line processing and tomographic tool}

\begin{figure*}[H]
\centering
\mbox{
\includegraphics[width=5cm]{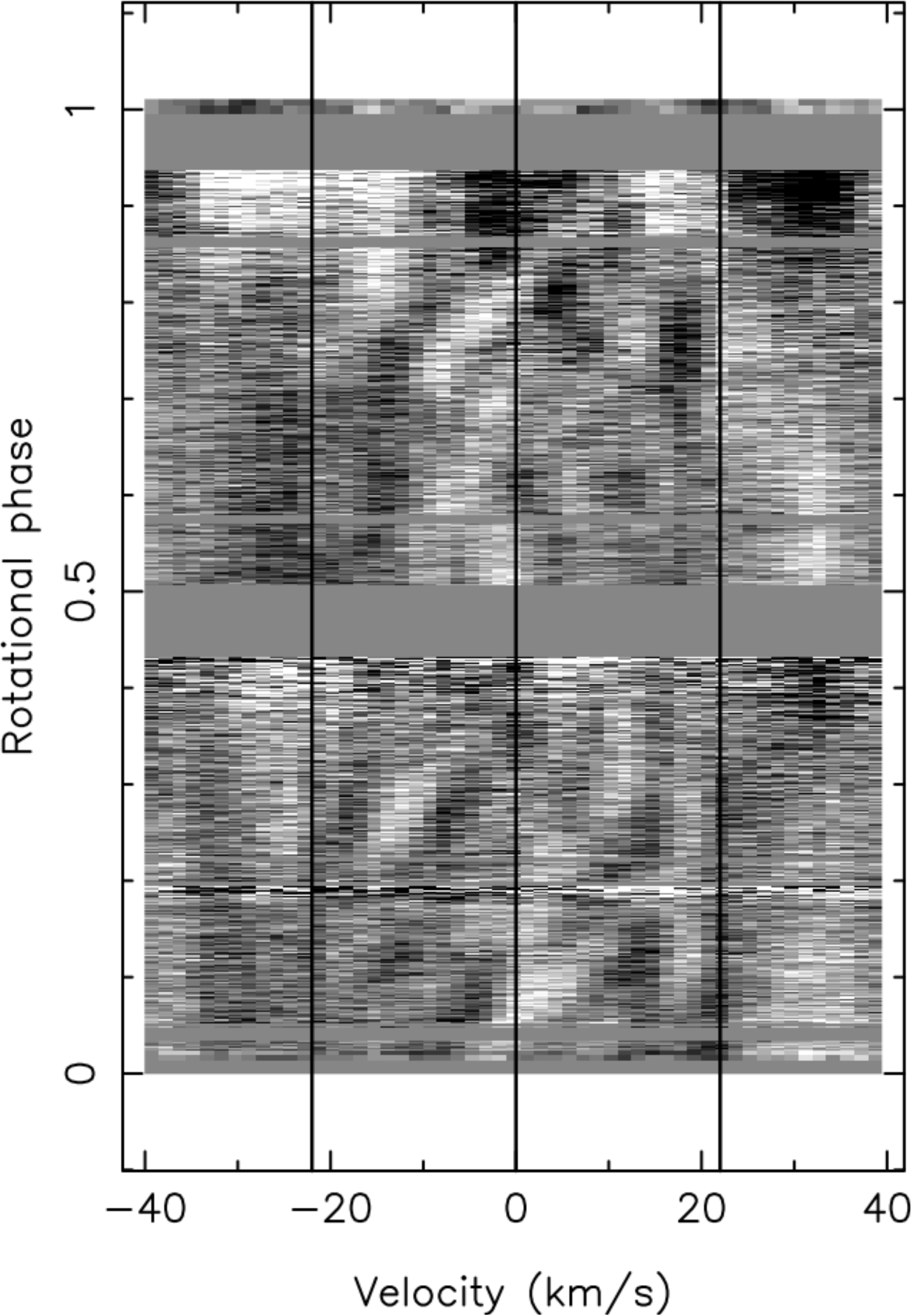}
\includegraphics[width=5cm]{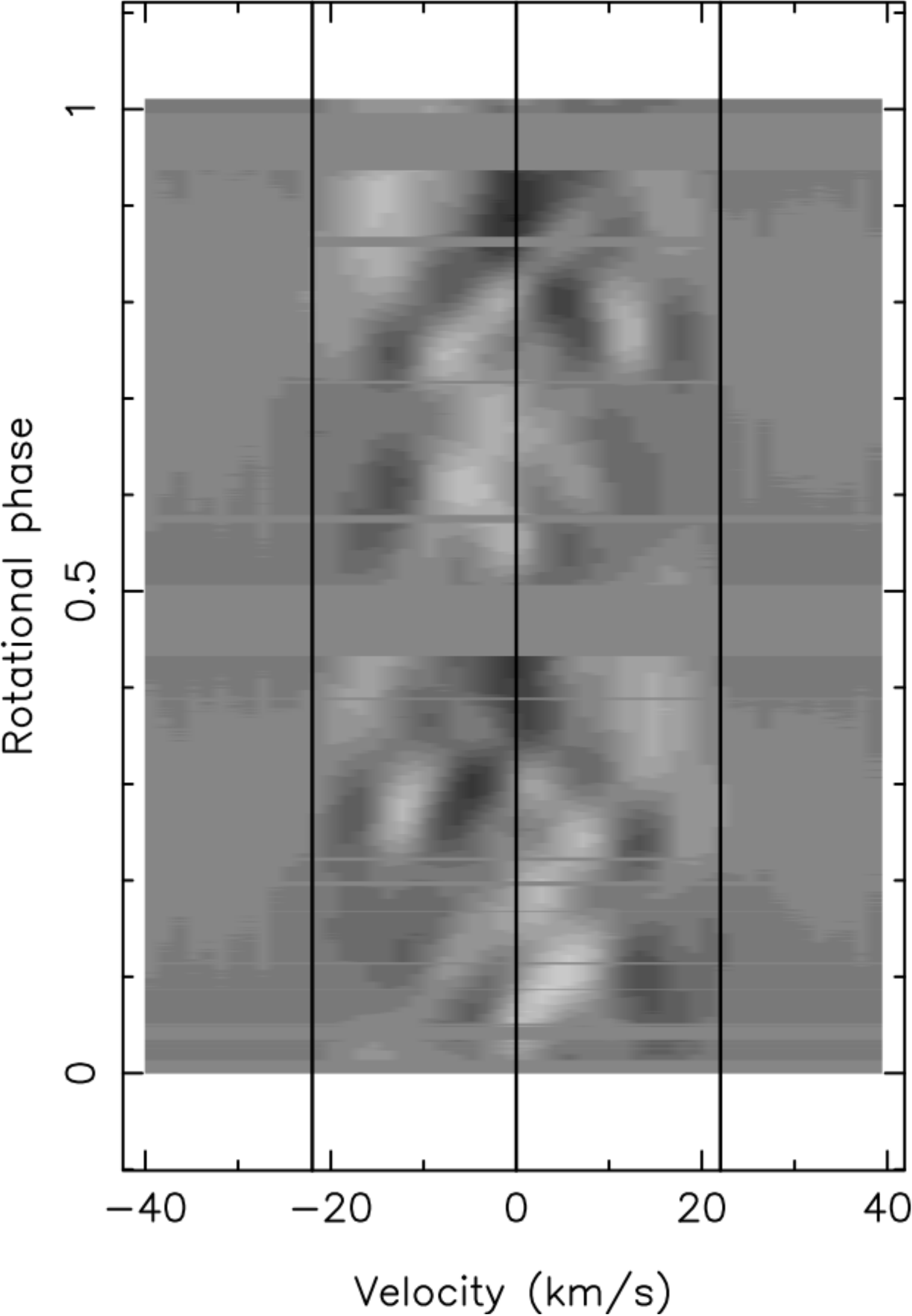}
\includegraphics[width=5cm]{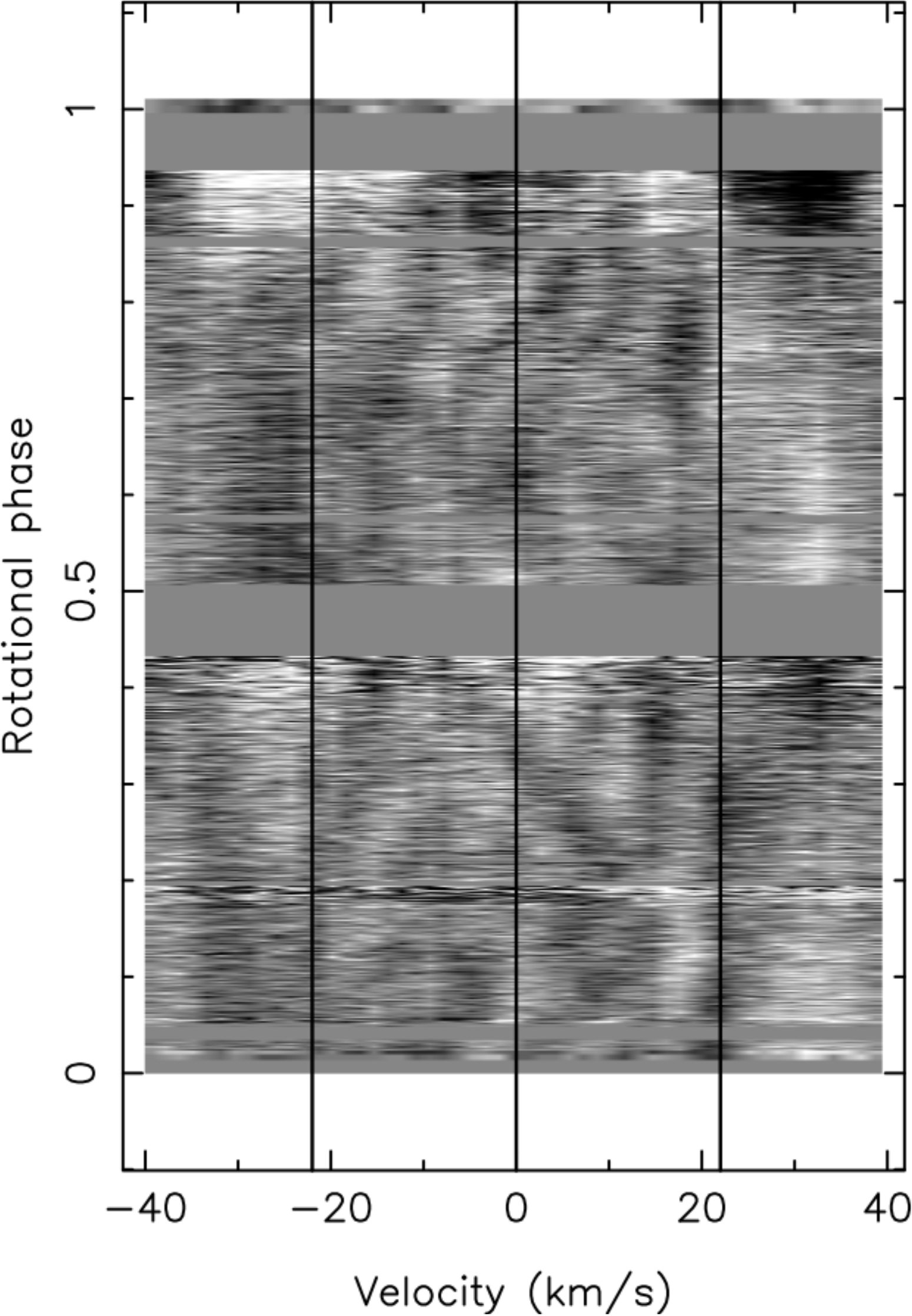}
}
\mbox{
\includegraphics[width=5cm]{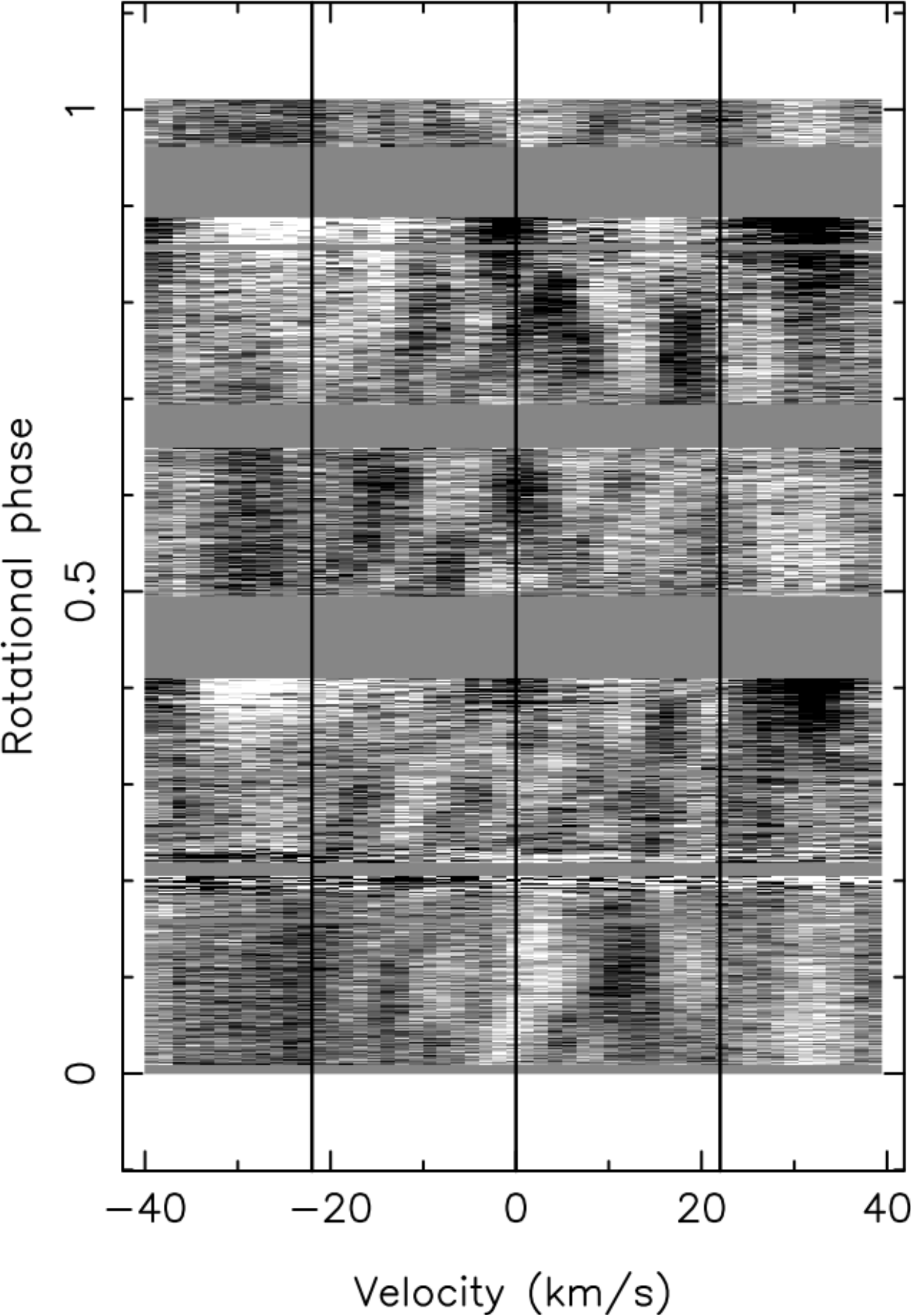}
\includegraphics[width=5cm]{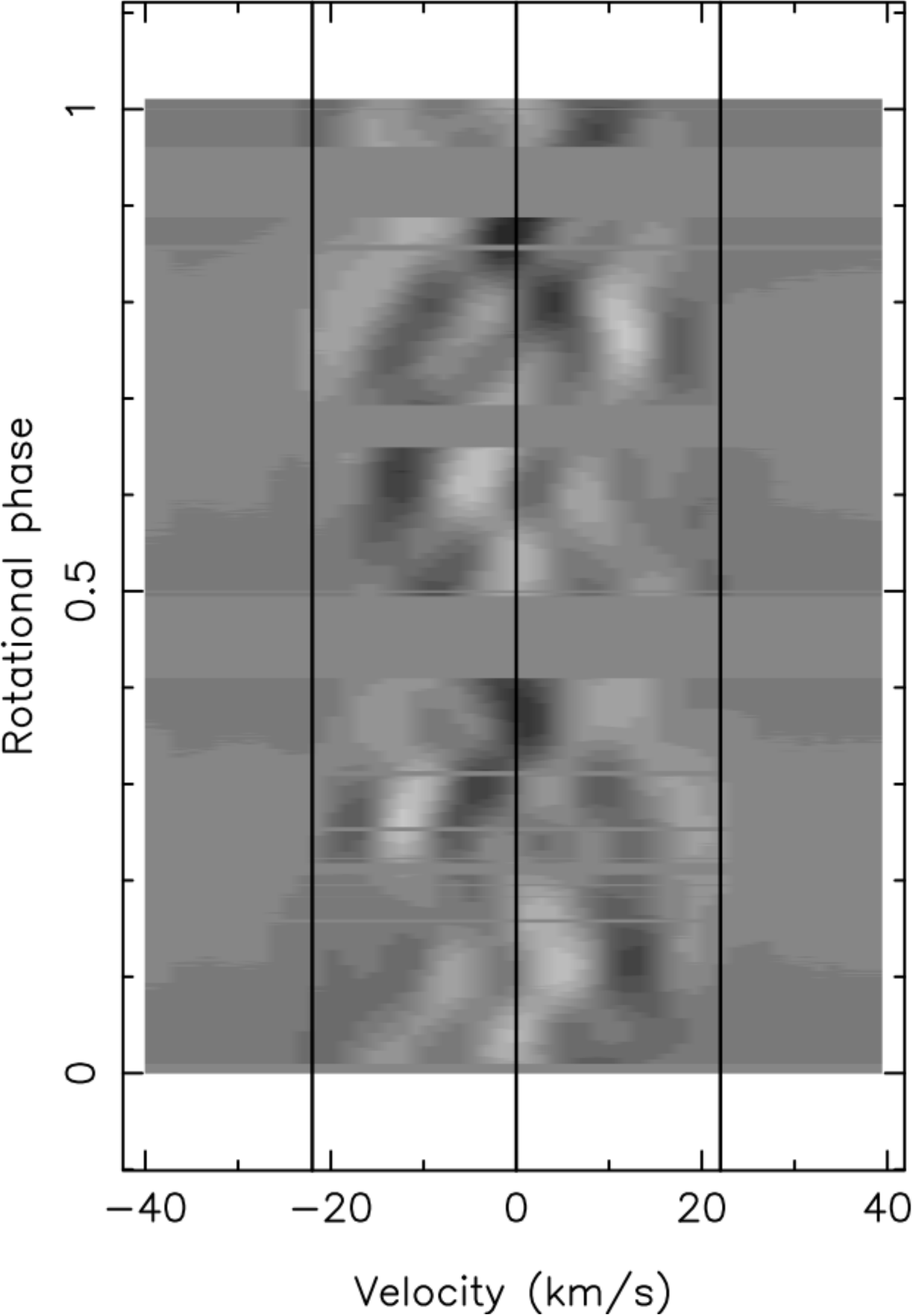}
\includegraphics[width=5cm]{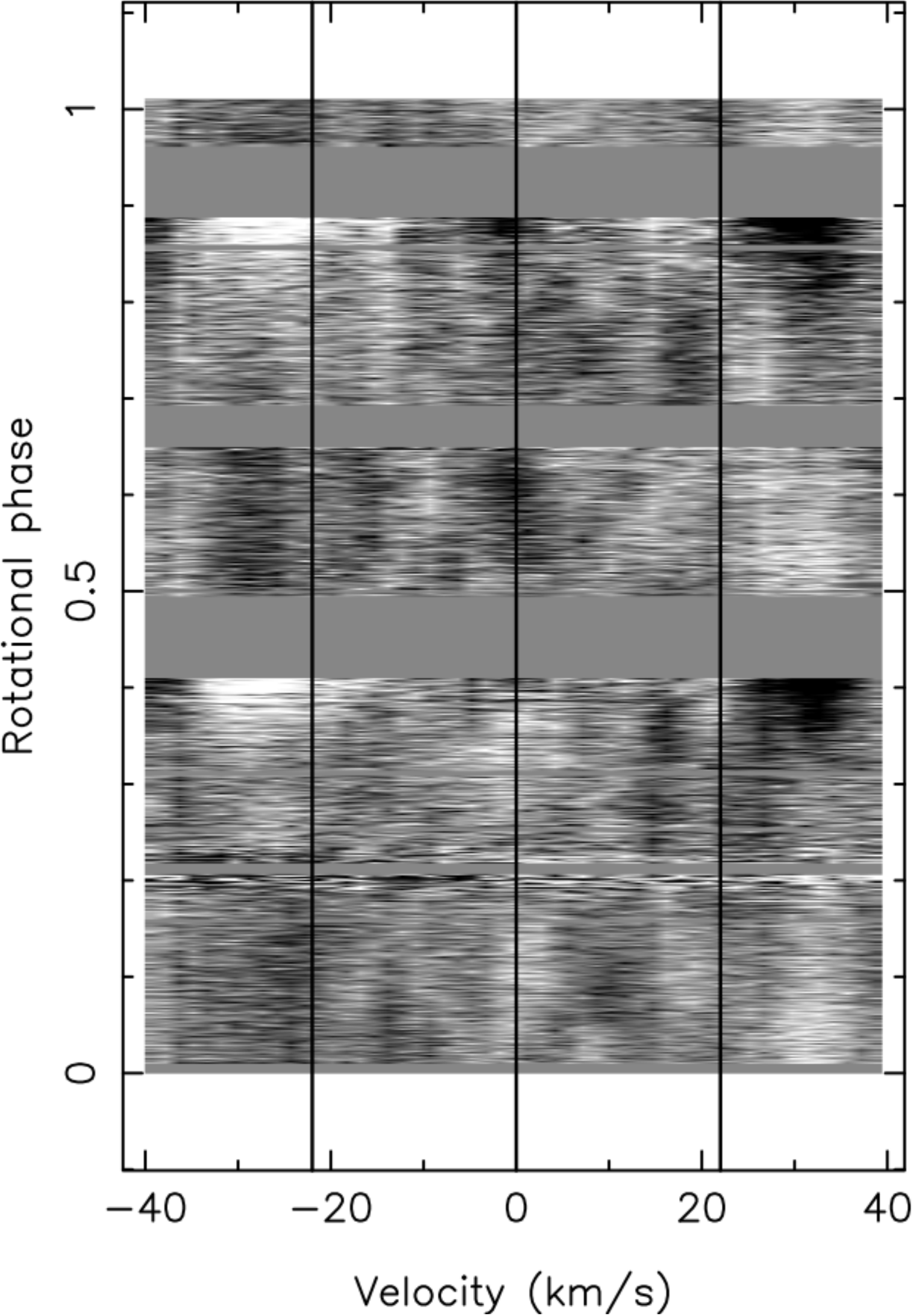}
}
\mbox{
\includegraphics[width=5cm]{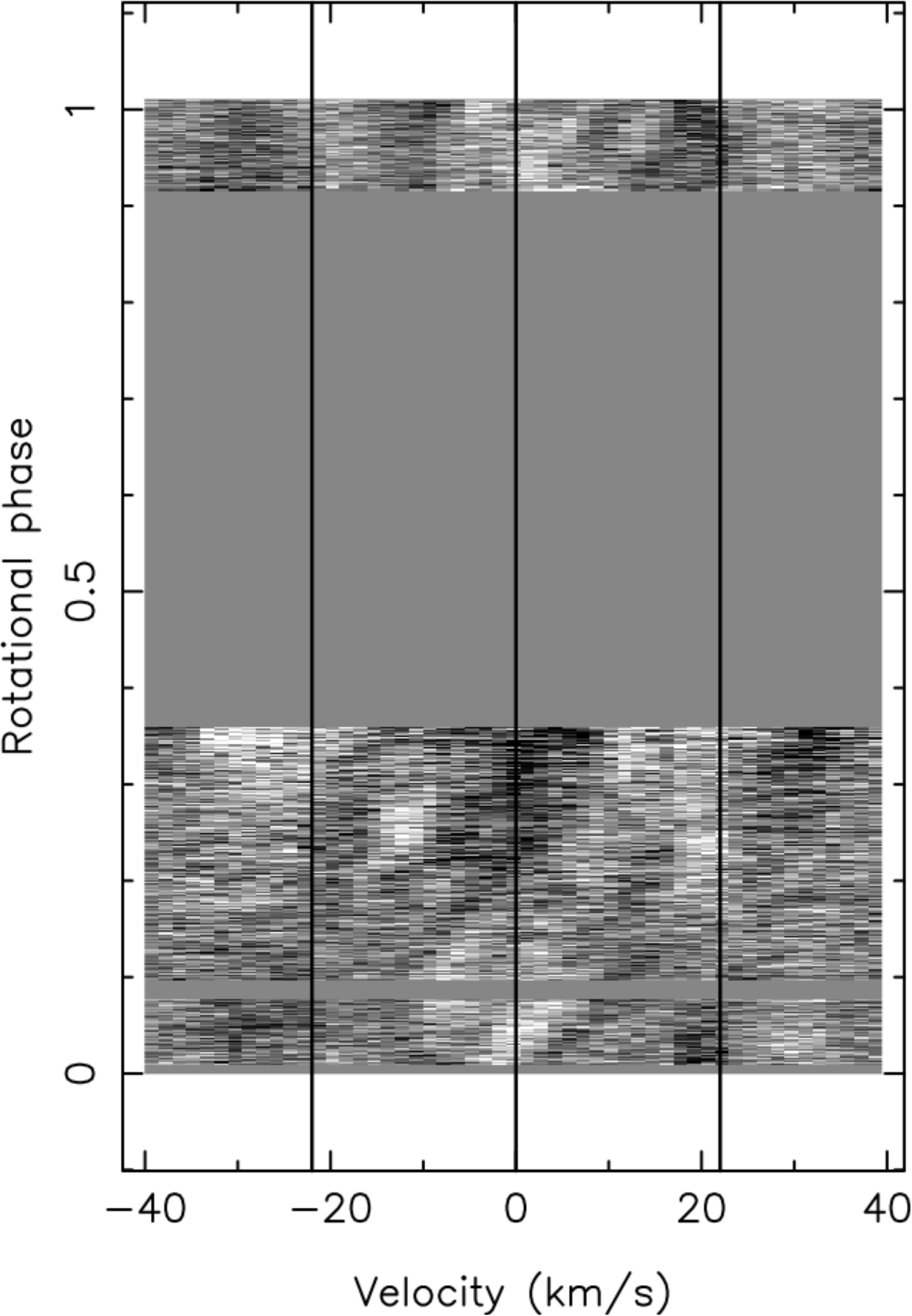}
\includegraphics[width=5cm]{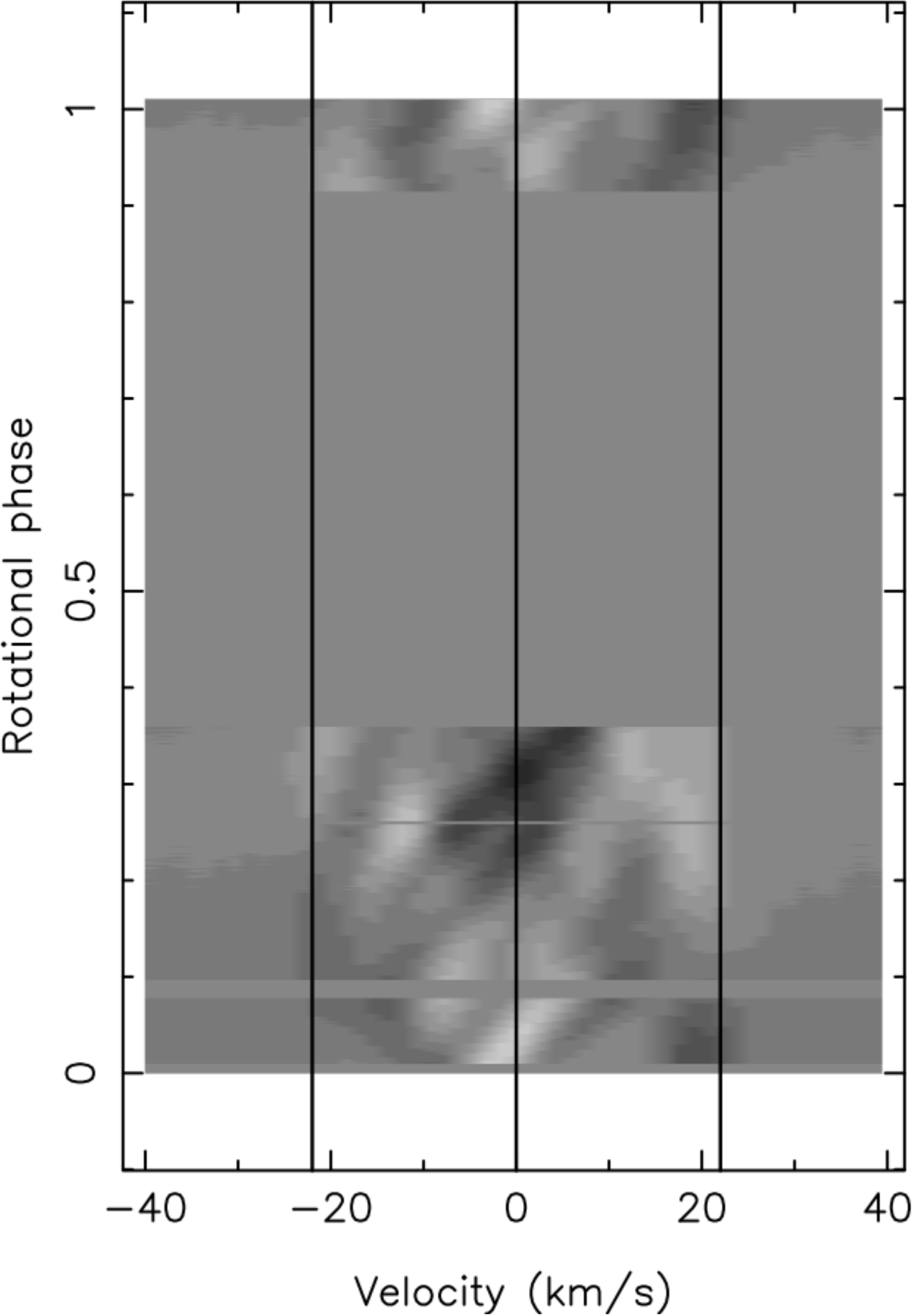}
\includegraphics[width=5cm]{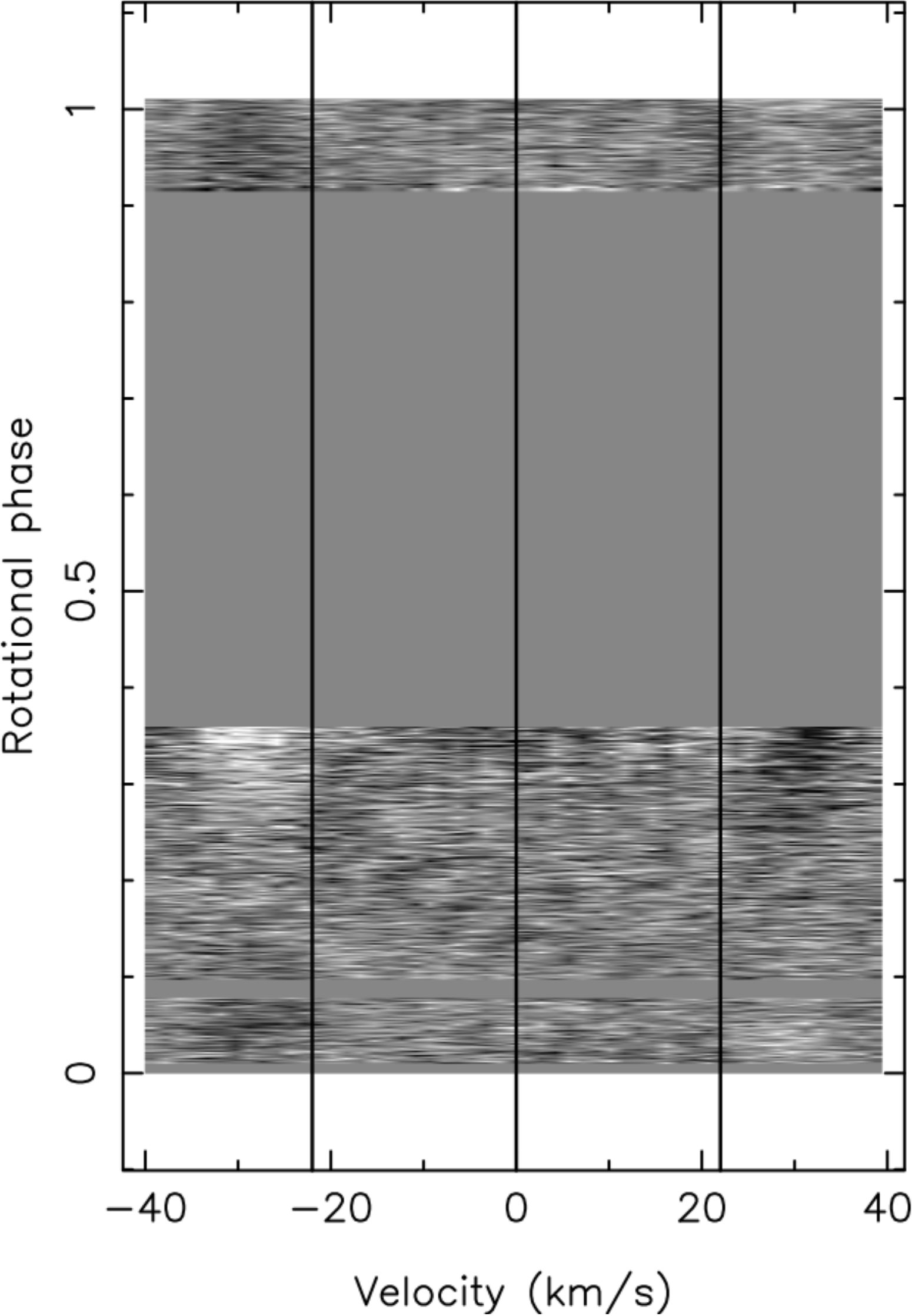}
}
\caption{LSD pseudo-line profiles of Vega represented as dynamic spectra, after subtraction of the phase-averaged profile. From top to bottom, we display the time sequences collected during nights 1-2, 3-4, and 5. The left, middle, and right columns show the observations, Doppler imaging models, and model residuals, respectively. The three vertical lines are plotted at line center, and at $\pm$~\vsin. The grey scale saturated at $\pm 0.045\%$.}
\label{fig:profiles}
\end{figure*}

The data set was obtained at the Haute Provence Observatory, using the SOPHIE \'echelle spectrograph \citep{bouchy06}, and was previously described in detail by \cite{boehm15}. The observations consist of a very dense temporal coverage of Vega during five consecutive nights in the summer of 2012, running from August 2 to August 6. Around 500 spectra were collected daily, for a total of 2,588 spectra during the whole campaign, with an average signal-to-noise ratio (\sn\ from now on) close to 850. Thanks to the 0.678~d rotation period of the target, about half a stellar rotation was densely sampled every night, and combining data from two consecutive nights offers a nearly complete phase coverage. For this reason, we chose to consider separately in our models three subsets constituted of data gathered during the first and second nights (nights 1-2 hereafter), the third and fourth nights (nights 3-4), and the fifth night (night 5), at the cost of a sparser phase coverage of our last subset. All observations were processed using the Least-Squares-Deconvolution method (LSD, \citealt{donati97,kochukhov10}), to produce a set of pseudo-line profiles with enhanced \sn. Details on the list of photospheric lines used for LSD can be found in \cite{boehm15}. The series of observed LSD profiles are plotted in the left panel of Fig. \ref{fig:profiles}.

Here we use this unique time-series to recover the distribution of surface inhomogeneities using tomographic modeling. The origin of the very small spectral signatures reported by \cite{boehm15} remain uncertain, the most obvious candidates being pulsations, chemical spots or temperature spots. Pulsations can easily be discarded, as they should leave their imprint at higher frequencies \citep{boehm15}. The hypothesis of chemical patches could, in principle, be tested by limiting the computation of LSD pseudo-profiles to spectral lines of individual chemical elements, checking if the signal is produced by specific species. Reducing the number of lines used for LSD would, however, increase the noise and significantly reduce the detectability of line profile variability. As a consequence, our data did not enable us to decide between chemical spots and temperature inhomogeneities. In the rest of the text, we assume that local temperature differences generate the spectral signatures. Whatever the actual nature of surface spots, our initial asumption will provide us with a reliable determination of their location and area, which is the prime objective of our study. From now on, and in spite of this uncertainty, we use the term "brightness spot" when referring to reconstructed surface features.

The code applied here is based on a maximum entropy approach, using a model of a spherical stellar surface divided into a grid of independent pixels. We employ a Unno line model based on a Voigt profile to describe local lines associated to each surface pixel, adopting a slope of the Planck function with continuum optical depth equal to 2.6, an opacity ratio equal to 2.5, a damping constant equal to 0.35, and Doppler broadening equal to 1.64 \kms\ (all these parameters were adjusted to fit Stokes I pseudo-profiles of Vega). The stellar model makes use of a linear limb darkening coefficient equal to 0.5. The code allows for both dark and bright spots to be reconstructed (see \citealt{donati16} and references therein). Doppler Imaging is especially difficult to apply to Vega, because the shape of LSD pseudo-line profiles is affected by the gravity darkening produced by the fast rotation \citep{takeda08}. This is possibly combined with an abnormal limb darkening related to the steep temperature gradient between the pole and equator. This complex situation is critical in view of the very small spectral features spotted in the dynamic spectra (where bumps and dips do not exceed $10^{-3}$ of the continuum level). As a consequence, inaccuracies in the line model may dominate over rotationally-modulated signatures. To circumvent this problem, we use the Doppler Imaging (DI) code detailed by \cite{hebrard15}, which is based on an original algorithm aimed at fitting the residuals to the average line profile, instead of the line profiles themselves (so that the exact shape of the average profile does not matter much). An important limitation of this approach is to ignore any axi-symmetric surface features, $e.g.$ any polar spot. This weakness is acceptable here, as the latitudinal dependence of surface brightness is expected to be dominated by gravity darkening. Another limitation of the DI code is the rough asumption of a spherical star, while the significant oblateness of Vega \citep{aufdenberg06} will lead the DI inversion to overestimate the latitude of spots. \vsin\ is kept equal to 22 \kms\ \citep{petit10}, but the rotation period was adjusted to the value of \cite{boehm15}. We stay with the 7\degr\ inclination angle proposed by \cite{takeda08} although slightly lower values were also suggested (down to 4.5\degr, see \citealt{yoon10} for a discussion), because our DI code is mostly insensitive to such small differences in this parameter value.

\section{Results}

\begin{figure}[H]
\centering
\includegraphics[width=6cm]{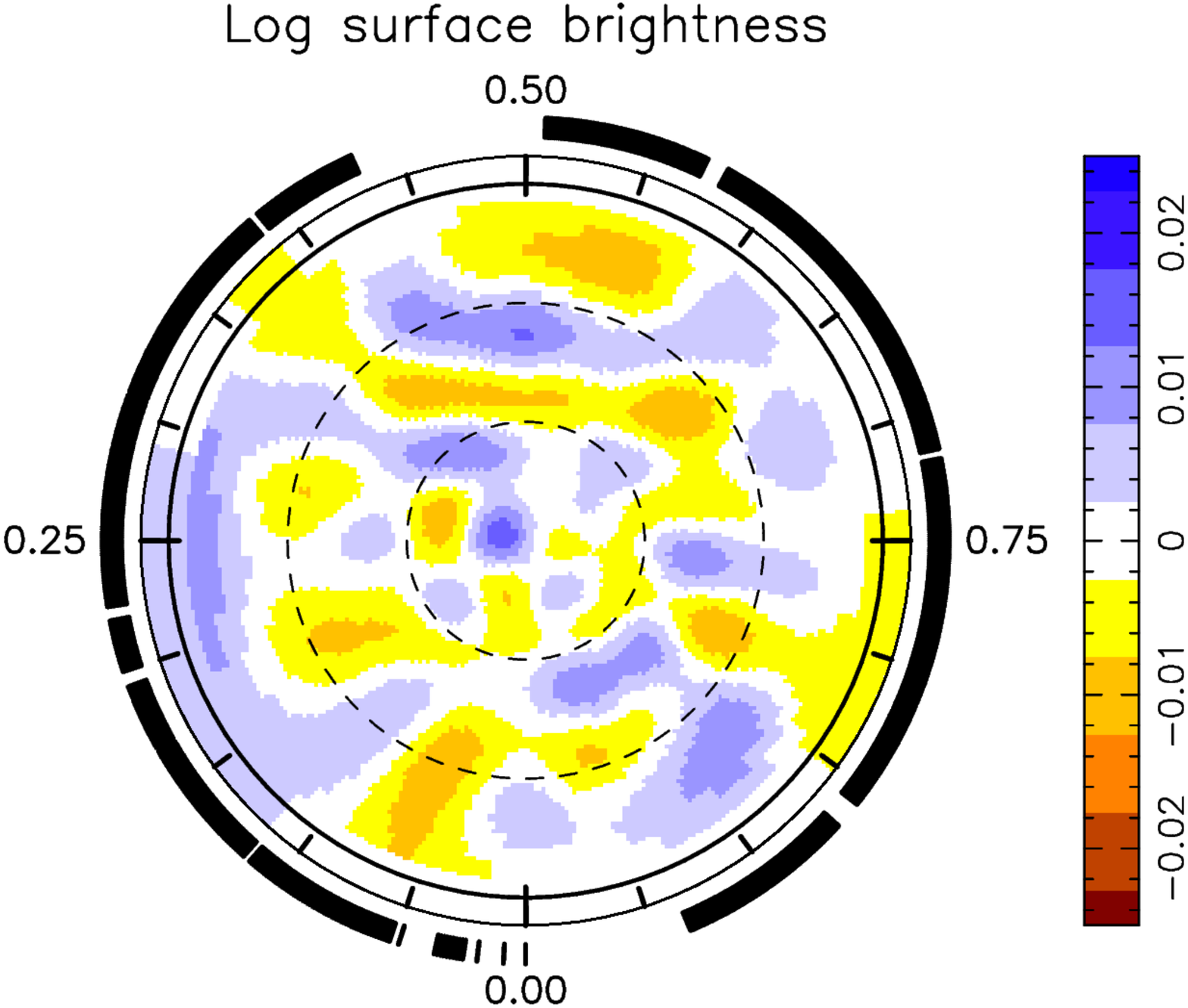}
\includegraphics[width=6cm]{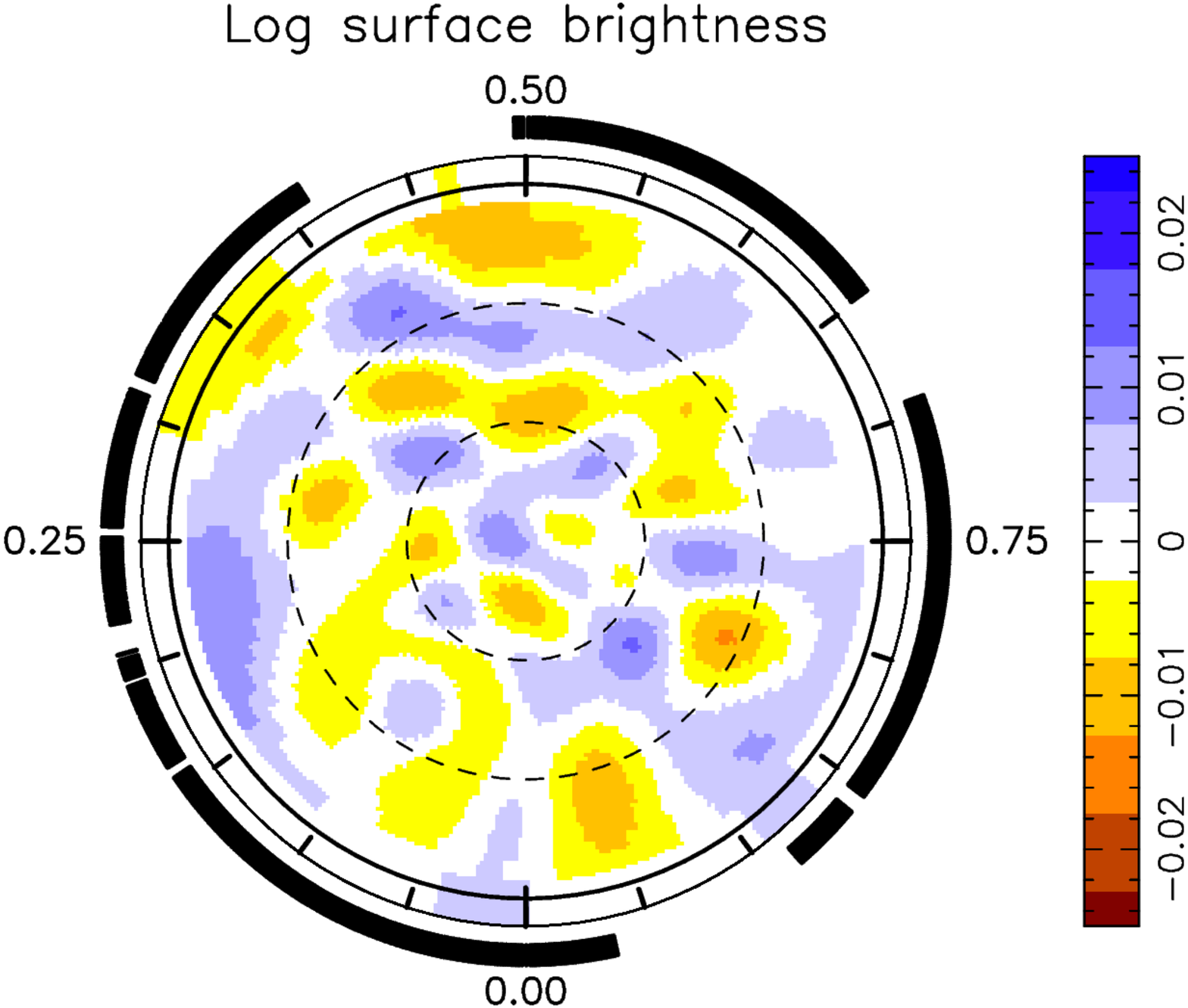}
\includegraphics[width=6cm]{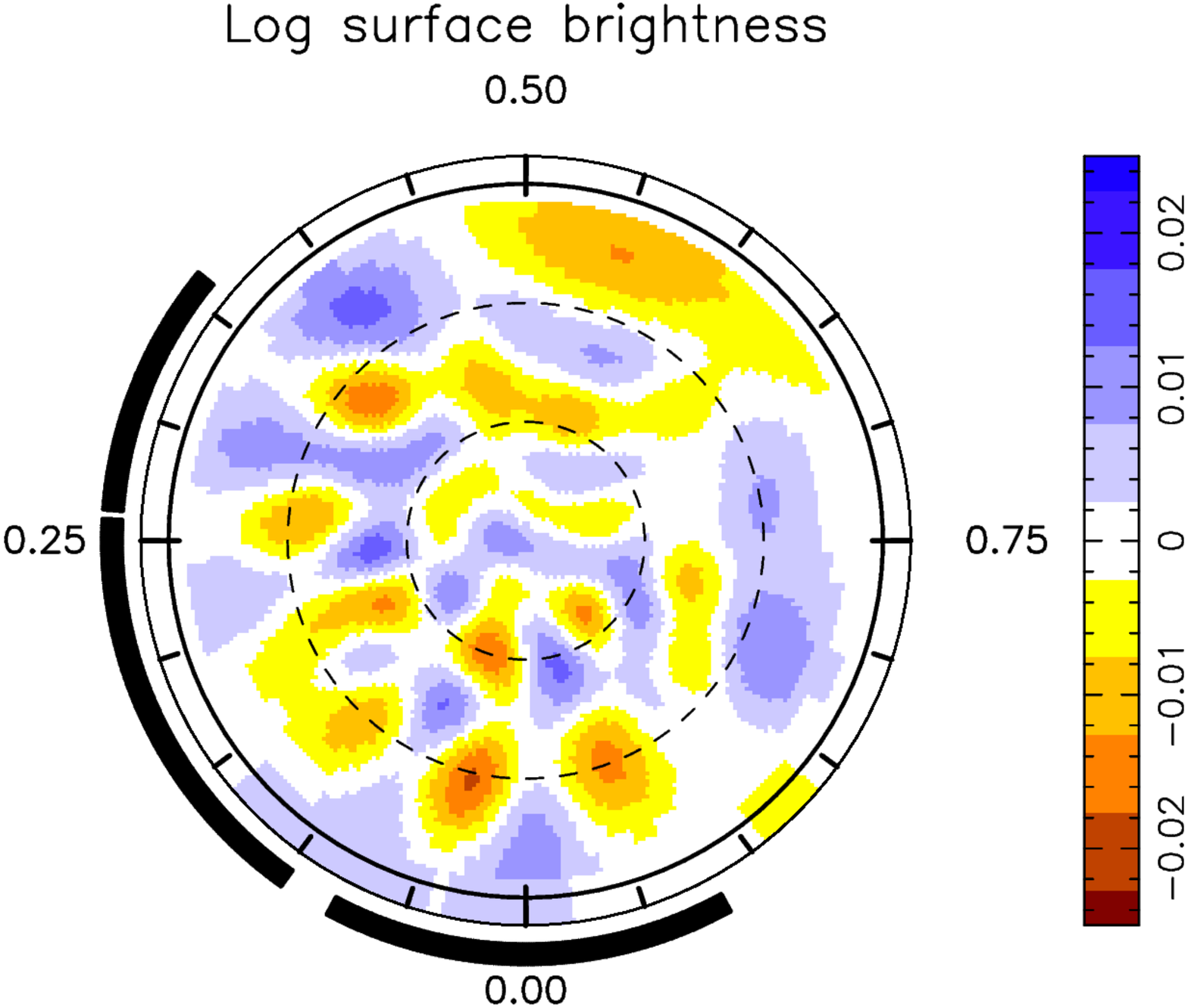}
\caption{Doppler maps obtained for the nights 1-2 (top), 3-4 (middle), and 5 (bottom). The visible stellar hemisphere is represented in a flattened polar projection, where the two dashed circles show parallels at latitude 30\degr\ and 60\degr, while the equator shows up as a full circle. The outer circle shows latitude -7\degr. The color scale represents the logarithmic, normalized brightness, with blue and yellow/brown patches depicting spots brighter and darker than the unspotted photosphere. The discontinuous black bands surrounding the maps illustrate the observed phases.}
\label{fig:maps}
\end{figure}

The brightness maps reconstructed from our three subsets are plotted in Fig. \ref{fig:maps}. The synthetic spectra produced by the DI code are illustrated in the middle panel of Fig. \ref{fig:profiles}, and the difference between the observations and the model are shown in the right panel of the same figure. 

\subsection{Dynamic spectra}
\label{sec:profiles}

From Fig. \ref{fig:profiles}, it can be noticed that most trails seen in the observed dynamic spectra are correctly duplicated in the synthetic time series produced by the DI code. With very little usable information in the line profiles, the \kis\ achieved by our model is lower than, but not very different from, the \kis\ of a model with no spots on stellar surface. The relative \kis\ difference between the two models ranges from 10\% to 25\%, depending on the map. While the percent change in \kis\ is modest, the improvement over a homogeneous model has a large statistical significance, due to the very large number of observed data points ($2.6\times 10^4$ to $5.4\times 10^4$ data points, depending on the data sub-set). Thanks to the absence of noise, the DI output dynamic spectra help to better distinguish these subtle trails. Most of them transit from the blue to the red wings of the profiles, as expected for a Doppler-shifted, rotationally-modulated signal. The observed mix of bright and dark strips are indicative of the presence of surface spots that can be darker or brighter than the average photospheric brightness (at a given latitude). Some of the trails go backwards ($e.g.$ the bright strip between phases 0.2 and 0.4 for nights 1-2), because the low inclination angle allows to follow some surface features in their transit behind the spin axis. The DI model remains close to the observations at phases where the continuum normalisation is not perfect or the noise level is slightly higher, presumaby because the spots are visible during most of the rotation period, limiting the influence of specific phases. 

The comparison of successive dynamic spectra highlights their mostly consistent content, but it is also possible to notice subtle differences. For instance, the dark strip crossing line center at phase 0.3 during night 5 is mostly absent from night 1-2 and 3-4. Another example is the bright backwards trail crossing line centre at phase 0.3 during nights 1-2 and 3-4, which is not seen at all during night 5. This simple observation suggests that some some level of surface evolution occurs within the limited time span of the five observing nights. 

\subsection{Brightness maps}

The brightness maps confirm most observations reported from the inspection of the dynamic spectra. They feature a complex pattern of bright and dark spots, with a roughly equal fraction of surface area brighter and darker than the average brightness at their latitude. Some spots are elongated in latitude, which is likely due to the intrinsically lower spatial resolution of DI codes in the latitudinal direction. This effect is maximal close to the equator, were the spot visibility is hampered by projection effects. Assuming that the smallest reconstructed spots are unresolved, so that their apparent area is directly related to the spatial resolution of the DI code, the latitudinal uncertainty in spot positioning can be estimated to be of about 30\degr\ in the worst cases (at low latitudes), while longitudinal positioning uncertainty is typically around 20\degr. These values increase when the phase coverage is poor, $e.g.$ between phases 0.35 and 0.9 of night 5, but spots above a latitude of about 40\degr\ seem to be less affected.

Comparing successive maps (based on independent data sets) helps to assess the reality of the reconstructed brightness structures. For instance, the dark spot seen at phase 0.05 and at a latitude of roughly 20\degr\ is consistently reconstructed from all subsets. Apparent modifications of its location, size and contrast may be related to differences in phase coverage, \sn\, or to occasional inaccuracies in continuum normalization. Spots reconstructed at low latitudes in the wide phase gap of night 5 have no obvious counterparts in nights 1-2 and 3-4, presumably because the DI model is poorly constrained for them. The same limitation does not affect higher latitudes, where the surface visibility is good during the whole stellar rotation. 

We also report differences between the maps that cannot be easily explained by differences in data quality. The dark spot reconstructed from nights 1-2 and 3-4 at phase 0.8, just above latitude 30\degr, is also seen in the dynamic spectrum (Fig. \ref{fig:profiles}), where its signature is the bright trail crossing the line center at phase 0.8, and the backward trail crossing line center at phase 0.3. This very recognizable spectral signature is consistently seen during nights 1-2 and 3-4. However, as stressed in Sect. \ref{sec:profiles}, the backward strip is absent from night 5. Accordingly, this spot is not seen on the surface map obtained for night 5. Another example is the extended bright spot only seen during night 5, at phase 0.4 and below latitude 30\degr, directly related to the dark strip that newly appeared in the dynamic spectrum of night 5, crossing line centre at phase 0.4.

\subsection{Cross-correlation maps}
\label{sec:crosscor}

Another way to search for temporal variability in a more global manner consists in cross-correlating the latitudinal rings of different surface maps, as proposed by \cite{donati97b}. Cross-correlation maps are very sensitive to map-to-map differences in the phase coverage \citep{petit02}, so that the brightness map of night 5 cannot be directly compared to the earlier subsets using this method. In this section, we replace this map by a new map (not shown here) reconstructed from nights 4-5, to ensure an even phase coverage for all maps.

\begin{figure}
\centering
\includegraphics[width=9cm]{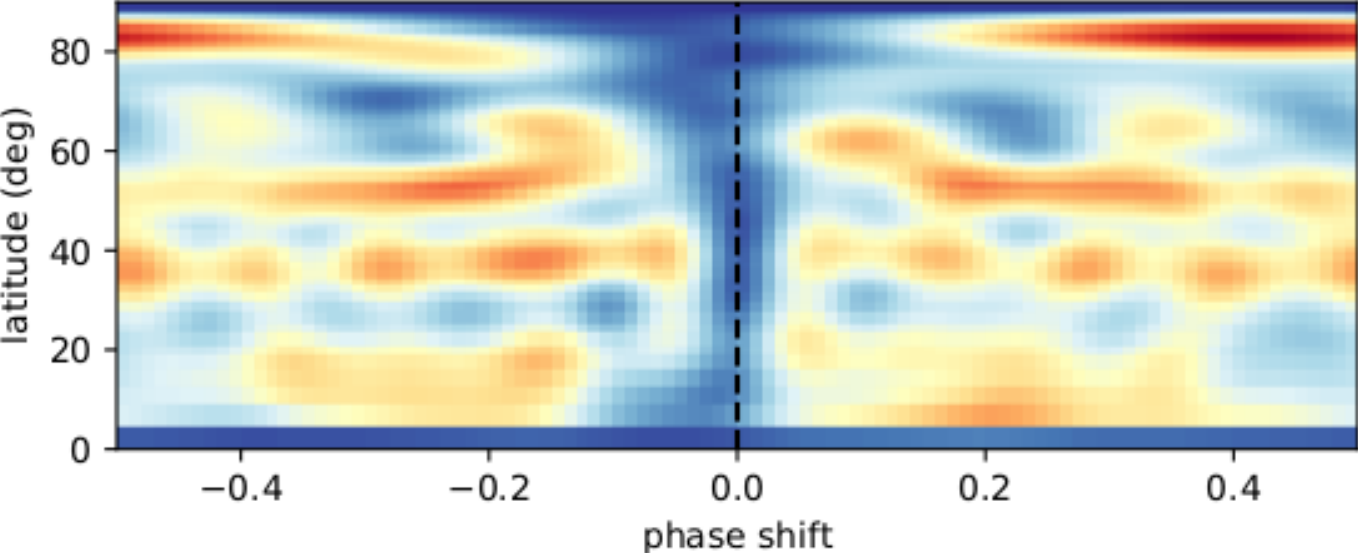}
\includegraphics[width=9cm]{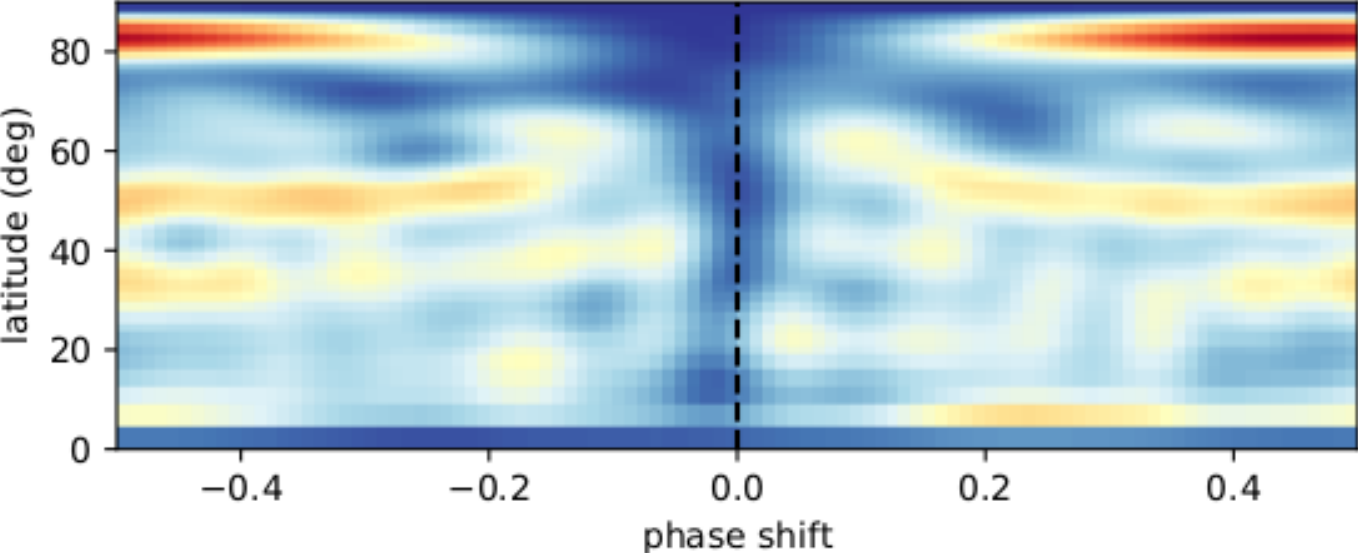}
\caption{Cross-correlation maps obtained from the comparison of nights 1-2 and 3-4 (upper panel), and nights 1-2 and 4-5 (lower panel). Bluer colors denote a higher correlation.}
\label{fig:cross}
\end{figure}

Two example cross-correlation maps are displayed in Fig. \ref{fig:cross}. In both maps, a clear correlation peak shows up, with a phase shift close to zero, confirming the good correspondance between the two spot distributions. However, the correlation peak displays some systematic latitudinal trends, with brightness features at the lowest latitudes that seem to be progressively displaced towards lower phases. A similar trend is also seen above latitude 50\degr, while mid-latitudes do not show any such phase shifts (while a small error in the adopted rotation period would produce a phase drift affecting all latitudes). 

We investigated the six possible combinations of two maps, chosen from the four maps reconstructed with nights 1-2, 2-3, 3-4 and 4-5. The outcome is generally consistent with the combinations illustrated in Fig. \ref{fig:cross}. At low latitudes, one noticeable exception is the comparison between nights 3-4 and 4-5, for which the shift is in the opposite direction. While all combinations display a consistent shift between latitudes 50\degr\ and 75\degr, the situation is more ambiguous above 75\degr, where one half of the cross-correlation maps show a drift with the opposite sign. As an alternate way to roughly estimate the significance of the phase shifts, we fitted gaussian functions on the cross-correlation peaks, and used their widths as a proxy of the error bar (following \citealt{donati97b}). By doing so, we obtained that the observed drifts reach, at best, the 1$\sigma$ level. 

We finally checked whether the phase shifts were increasing when we increase the time gap between the two brightness maps to be compared. We find marginal evidence of such effect, although the differences remain below 1$\sigma$, and some map combinations are not consistent with this view.    

Although not fully conclusive, these tests suggests that latitudinal differential rotation may contribute to the modification of the spot distribution, and that this surface shear may take the form of zonal flows confined in two belts rotating faster than the rest of the stellar surface, instead of the smooth latitudinal dependence observed on Sun-like stars \citep{barnes05}. 


\section{Conclusions}

An intensive monitoring of Vega during five consecutive nights revealed a complex spot pattern, featuring a mix of dark and bright regions spread over the whole visible part of the stellar surface. A rapid spot evolution is observed in the dynamic spectra, in the visual comparison of individual brightness maps, and in cross-correlation maps. The photospheric dynamics are seen as both local modifications of the spot distribution (spot emergence and decay), and possibly as zonal flows affecting the equatorial region and a high latitude belt, which both appear to rotate faster than the rest of the surface. This is the first direct evidence of a fast evolving spot pattern on an intermediate mass star. A long-term evolution of chemical clouds was previously reported for the HgMn star $\alpha$~Andromedae \citep{kochukhov07}, but over a much longer timescale of seven years. Future observations, covering a larger time-span, may help to confirm and study in greater details this progressive surface distortion.    

It is tempting to compare the spot patterns presented here to the complex magnetic geometry reported by \cite{petit10}. However, given the observed level of surface brightness variability, the three year gap separating the two data sets makes the comparison difficult. We checked that the cross-correlation of the spot maps with the magnetic maps does not show any significant correlation peak. However, any possible counterpart of the polar magnetic region reported by \cite{petit10} cannot appear in our brightness maps, owing to a DI algorithm that ignores all axisymmetric structures. Only future, strictly simultaneous data will help test the possible correspondance of brightness and magnetic patches, and bring more observational clues to understand the origin of Vega-like magnetism.    






\bibliographystyle{mnras}
\bibliography{vega} 

\bsp	
\label{lastpage}
\end{document}